\documentclass[11pt,letterpaper]{article}

\usepackage[vmargin=1in]{geometry}
\usepackage[T1]{fontenc}
\usepackage{lmodern,microtype}
\usepackage{amsmath,amssymb,bm}
\usepackage{graphicx,xcolor,tikz}
\usetikzlibrary{arrows.meta,calc,positioning}
\usepackage[font=small,labelfont=bf]{caption}
\usepackage[numbers]{natbib}
\usepackage{hyperref}

\newtheorem{remark}{Remark}

\definecolor{stretch}{RGB}{181,55,55}
\definecolor{squeeze}{RGB}{43,112,168}
\definecolor{earth}{RGB}{47,114,83}
\newcommand{\rr}{\mathbf r}
\newcommand{\vv}{\mathbf v}
\newcommand{\RR}{\mathbf R}
\newcommand{\GG}{\mathbf G}
\renewcommand{\gg}{\mathbf g}
\newcommand{\TT}{\boldsymbol{\mathsf T}}
\newcommand{\CC}{\boldsymbol{\mathsf C}}
\renewcommand{\tt}{\boldsymbol{\eta}}
\newcommand{\xxi}{\boldsymbol\xi}
\newcommand{\Id}{\boldsymbol{\mathsf I}}
\newcommand{\dd}{\mathrm d}
\newcommand{\transpose}{\top}
\title{Tendex Lines and Short-Time Deformation: The Geometry of Lunar Tides}

\author{Francisco J. Beron-Vera\\
Department of Atmospheric Sciences\\
Rosenstiel School of Marine, Atmospheric, and Earth Science\\
University of Miami\\
Miami, FL 33149\\
\href{mailto:fberon@miami.edu}{\texttt{fberon@miami.edu}}
}
\date{November 9, 2026}

\begin{document}
\maketitle

\begin{abstract}
    Tendex lines and the stretchlines and squeezelines of Lagrangian coherent structure theory share a geometric construction: they integrate eigenvector fields of symmetric tensors. We connect their dynamical interpretations through the short-time deformation of particles released with zero initial velocity relative to a freely falling reference in an autonomous gravitational field. The leading departure of the Cauchy--Green tensor from the identity is quadratic in time and equals the tidal tensor multiplied by the squared observation time. The tidal tensorlines therefore give the limiting principal stretching and squeezing directions. For the planar lunar field, these tensorlines are radial lines and concentric circles centered on the Moon. The same tidal tensor defines a Lorentzian metric whose null directions have no quadratic change in line-element length. Its null geodesics are logarithmic spirals, describing the short-time limit of finite-time unit-stretch geometry. Near Earth, the principal tensorlines clarify the two outward tidal tendencies, while the null directions distinguish vanishing leading extension from vanishing relative acceleration.
\end{abstract}

\noindent\textbf{Curves that reveal tidal stretching near a gravitating body and curves that reveal stretching in a fluid flow are usually discussed in different settings. Both follow principal directions of symmetric tensor fields. For particles released at rest relative to a freely falling reference, this geometric resemblance has a direct dynamical expression: the tidal tensor determines their leading short-time deformation. Earth and nearby particles are subject to the Moon's gravitational field. Relative to Earth's freely falling center, longitudinal stretching has two outward ends, clarifying why attraction toward the Moon produces an outward tidal tendency on both the near and far sides. The same tensor identifies directions with no leading change of length, which integrate to logarithmic spirals. Material segments initially aligned with these directions turn even though their leading extension vanishes, distinguishing relative acceleration, displacement, and deformation.}

\section{Introduction}

Tensor eigenvector curves provide a geometric description of deformation in both gravitational mechanics and fluid dynamics. Tendex lines integrate the principal directions of a gravitational tidal tensor \citep{NicholsEtAl2011,Thorne2014}, while stretchlines and squeezelines in Lagrangian coherent structure (LCS) theory integrate those of a finite-time Cauchy--Green tensor \citep{Haller2023}. The former describe relative acceleration; the latter describe accumulated stretching. We ask how these two tensorline constructions are related dynamically.

We consider a smooth family of particles released with zero initial velocity relative to a freely falling reference in an autonomous gravitational field. For a short forward time $\tau$, the Cauchy--Green tensor satisfies $\CC_\tau(\rr_0)=\Id+\tau^2\TT(\rr_0)+O(\tau^4)$, where $\TT(\rr_0)$ is the tidal tensor at the initial position. Wherever the tidal eigenvalues are distinct, the corresponding principal directions are therefore the limits of the finite-time stretching and squeezing directions. We derive this relation from the particle dynamics and establish the role of the initial velocity prescription. The result describes leading material deformation without, by itself, identifying tidal tensorlines as coherent structures. To the best of our knowledge, this dynamical connection between gravitational tendex lines and the limiting Cauchy--Green stretchlines and squeezelines of particles released at relative rest has not previously been identified. In the planar lunar example, the indefinite tidal tensor also defines a Lorentzian metric. Its null directions have vanishing quadratic extension and integrate explicitly to logarithmic spirals, describing the short-time limit of finite-time unit-stretch geometry.

The classical lunar field makes these connections explicit. A frame following Earth's freely falling center separates common acceleration from the differences that deform a particle configuration. The mechanical foundation is established: Newton's Corollary VI to the laws of motion removes common acceleration from relative motion, and Book I, Proposition LXVI, Corollary XIX applies unequal attraction to tides \citep{Newton1999}. Oceanographic treatments develop the corresponding tide-generating force and potential \citep{Hendershott2005}; a free-fall formulation is given by \citet[Sec.~1.1]{CreightonAnderson2011}. Here the integrated tensorline field reveals the deformation geometry. Near Earth, longitudinal stretching produces outward relative accelerations on opposite sides of the center, accompanied by transverse squeezing. The null directions further distinguish vanishing leading extension from vanishing relative acceleration. This interpretation concerns tidal forcing and short-time deformation; calculating sea-surface elevation requires the ocean's response as well.

Throughout, free fall means motion under gravity alone and imposes no restriction on initial velocity. Acceleration determines the change in velocity, not the direction of motion. Our release prescription sets the particles' initial velocities to zero relative to the freely falling reference, whose own velocity need not vanish. Their relative displacements then follow their initial relative accelerations to leading quadratic order in time. Subsequent statements connecting acceleration directions to initial stretching, squeezing, or displacement use this prescription; subtracting the reference's acceleration does not require it.

\section{Relative acceleration in a freely falling frame}

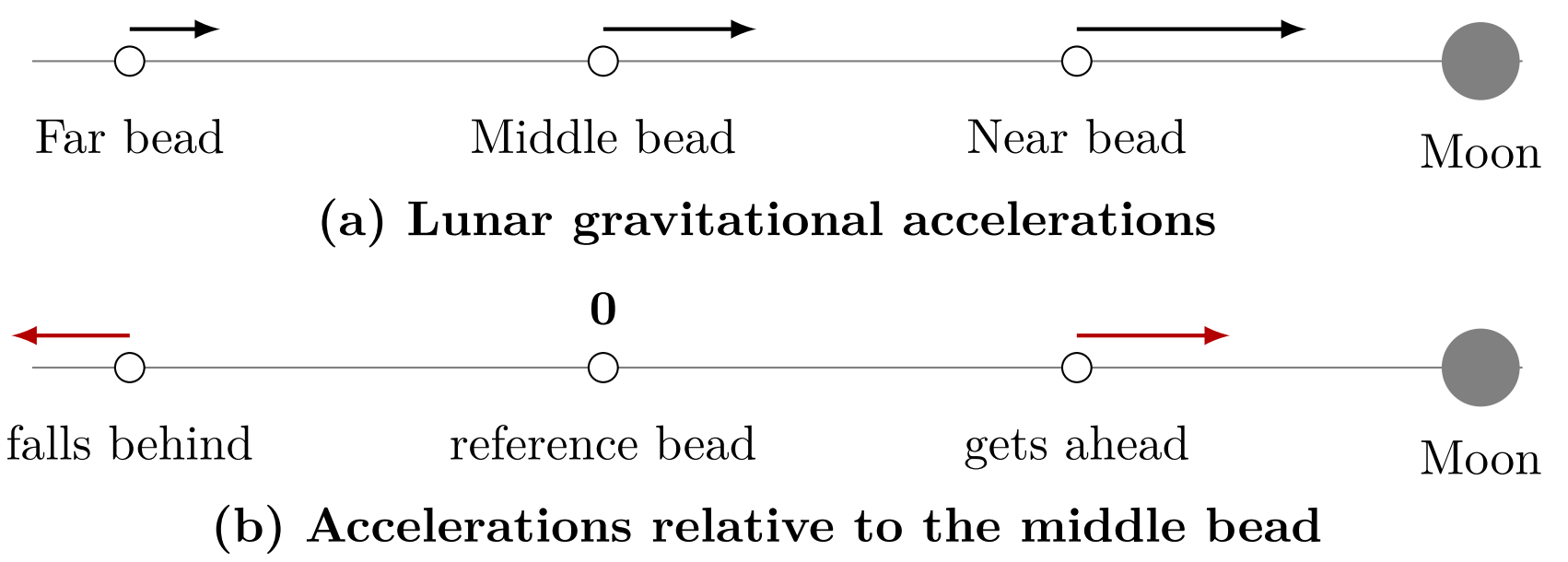
\begin{figure}[htbp]
    \centering
    \begin{tikzpicture}[x=1cm,y=1cm,>=latex,font=\small]
        \draw[gray] (-.1,1.8)--(10.6,1.8);

        \foreach \x in {0.6,4,7.4}{
            \filldraw[fill=white,draw=black]
                (\x,1.8) circle (.105);
        }

        \draw[->,thick] (.6,2.03)--(1.25,2.03);
        \draw[->,thick] (4,2.03)--(5.1,2.03);
        \draw[->,thick] (7.4,2.03)--(9.05,2.03);

        \node[below=8pt] at (.6,1.8) {Far bead};
        \node[below=8pt] at (4,1.8) {Middle bead};
        \node[below=8pt] at (7.4,1.8) {Near bead};

        \fill[gray] (10.3,1.8) circle (.28);
        \node[below=11pt] at (10.3,1.8) {Moon};

        \node[anchor=north,align=center] at (5.175,.95)
            {\textbf{(a) Lunar gravitational accelerations}};

        \draw[gray] (-.1,-.4)--(10.6,-.4);

        \foreach \x in {0.6,4,7.4}{
            \filldraw[fill=white,draw=black]
                (\x,-.4) circle (.105);
        }

        \draw[->,red!70!black,thick]
            (.6,-.17)--(-.25,-.17);
        \draw[->,red!70!black,thick]
            (7.4,-.17)--(8.5,-.17);

        \node[above=5pt] at (4,-.4) {$\mathbf{0}$};
        \node[below=8pt] at (.6,-.4) {falls behind};
        \node[below=8pt] at (4,-.4) {reference bead};
        \node[below=8pt] at (7.4,-.4) {gets ahead};

        \fill[gray] (10.3,-.4) circle (.28);
        \node[below=11pt] at (10.3,-.4) {Moon};

        \node[anchor=north,align=center] at (5.175,-1.25)
            {\textbf{(b) Accelerations relative to the middle bead}};
    \end{tikzpicture}
    \caption{One attraction, two outward relative accelerations. (a) All three beads accelerate toward the Moon, with acceleration increasing from the far bead to the near bead. (b) Subtracting the middle bead's acceleration gives outward relative accelerations on both sides. The leftward arrow represents weaker attraction toward the Moon, not repulsion. Arrow lengths are schematic.}
    \label{fig:beads}
\end{figure}

Consider three unconnected test beads on the Earth--Moon axis, with the Moon to their right. The middle bead represents Earth's center. All three accelerate toward the Moon, but the near bead accelerates most and the far bead least. Released with equal velocities, the near bead moves ahead of the middle bead and the far bead falls behind it. Their separations from the middle bead initially increase on both sides (Fig.~\ref{fig:beads}). Beads displaced transversely instead have relative accelerations toward the Earth--Moon axis. Together these motions give longitudinal stretching and transverse squeezing.

The beads experience lunar gravity alone; their mutual attraction is neglected. For Earth's matter, this isolates the lunar contribution by setting aside Earth's self-gravity, pressure, and material stresses. The Sun is also omitted. We use axes that translate with the reference bead without rotating. Thus no centrifugal or Coriolis terms arise from a rotation of the coordinates. This construction describes freely falling probes of the field, rather than the full dynamical response of the ocean.

A freely falling configuration needs no supporting force to share the reference's gravitational acceleration \citep[Sec.~1.1]{CreightonAnderson2011}. In a nonrotating frame translating with that reference, a spatially uniform gravitational field would therefore produce no relative acceleration. In a nonuniform field, the same change of frame removes the acceleration at the origin but leaves the differences across the configuration. Free fall thus eliminates neither gravity nor its tidal effects; it separates common acceleration from the relative accelerations that can deform an extended configuration. The same distinction underlies the absence of supporting forces for an astronaut inside an orbiting space station or a person inside an elevator freely falling in vacuum.

We work in the plane $\mathbb R^2$. At a chosen instant place Earth's center at $\mathbf0$ and the Moon at $\RR=(d,0)$, and write $\rr=(x,y)\in\mathbb R^2$. If the Moon has mass $M$, its force on a test mass $m$ is $\GG(\rr)=m\gg(\rr)$, where $\gg(\rr)=-GM(\rr-\RR)/|\rr-\RR|^3$. This is the restriction of the three-dimensional Newtonian field to the plane, with $|\cdot|$ denoting the Euclidean norm. The acceleration of the reference bead is $\gg(\mathbf0)$, so the relative equation is
\begin{equation}
    \ddot\rr=\gg(\rr)-\gg(\mathbf0)
    =GM\left(
    \frac{(d-x,-y)}{[(d-x)^2+y^2]^{3/2}}-\frac{(1,0)}{d^2}
    \right).
    \label{eq:relative}
\end{equation}
This subtraction removes the reference bead's acceleration while retaining the differences across the particle configuration. We freeze the orbital geometry by holding $\RR=(d,0)$ fixed in the translating frame. The resulting gravitational field has no explicit time dependence, so Eq.~\eqref{eq:relative} is autonomous.

A second subtraction compares two particles at $\rr$ and $\rr+\delta\rr$. Their separation obeys $\delta\ddot\rr=\gg(\rr+\delta\rr)-\gg(\rr)$: the reference acceleration $\gg(\mathbf0)$ cancels. For infinitesimal separations this gives the variational equation
\begin{equation}
    \delta\ddot\rr=\TT(\rr)\,\delta\rr,
    \qquad \TT=D\gg,
    \label{eq:variation}
\end{equation}
where $\TT$ is the planar tidal tensor, the Jacobian of the lunar gravitational acceleration field. It is evaluated at the evolving reference particle position $\rr(t)$. This linearization describes neighboring motions and does not require an equilibrium. It is distinct from replacing the tidal tensor throughout Earth's neighborhood by its value at the center. Common acceleration can therefore be removed without removing tidal deformation across an extended configuration, even though its individual particles are freely falling.

\section{The geometric connection: lunar tidal tensorlines}

Let $\mathbf s=\rr-\RR$ and $s=|\mathbf s|$. Differentiation gives the symmetric tensor
\begin{equation}
    \TT(\rr)=\frac{GM}{s^3}
    \left(3\frac{\mathbf s\mathbf s^\transpose}{s^2}-\Id\right),
    \qquad s^2=(x-d)^2+y^2.
    \label{eq:tensor}
\end{equation}
This is the exact planar restriction of the three-dimensional point-mass field, not a two-dimensional gravitational law. In increasing eigenvalue order, its orthonormal eigenpairs are
\begin{equation}
    \begin{aligned}
        \lambda_1&=-\frac{GM}{s^3},
        &\xxi_1&=\frac1s\begin{pmatrix}y\\d-x\end{pmatrix},\\[3pt]
        \lambda_2&=\frac{2GM}{s^3},
        &\xxi_2&=\frac1s\begin{pmatrix}d-x\\-y\end{pmatrix}.
    \end{aligned}
    \label{eq:eigenpairs}
\end{equation}
A separation aligned with $\xxi_2$ has outward relative acceleration; one aligned with $\xxi_1$ has inward relative acceleration. These are stretching and squeezing tendencies: an instantaneous change of length also depends on the particles' relative velocity. The omitted spatial direction has a second squeezing eigenvalue $-GM/s^3$, so the full three-dimensional tensor is trace-free, whereas its planar restriction has trace $GM/s^3$.

These principal directions define the planar classical tendex field of a point mass \citep{NicholsEtAl2011}. Their convention uses $\boldsymbol{\mathcal E}=-\TT$ and $\ddot{\delta\rr}=-\boldsymbol{\mathcal E}\delta\rr$, so negative tendicity denotes stretching, whereas our positive $\lambda_2$ does so. The sign reversal leaves the unoriented eigenvector curves unchanged. The name \emph{tendex}, coined by David Nichols from the Latin \emph{tendere}, to stretch, reflects this physical interpretation \citep{NicholsEtAl2011,Thorne2014}.

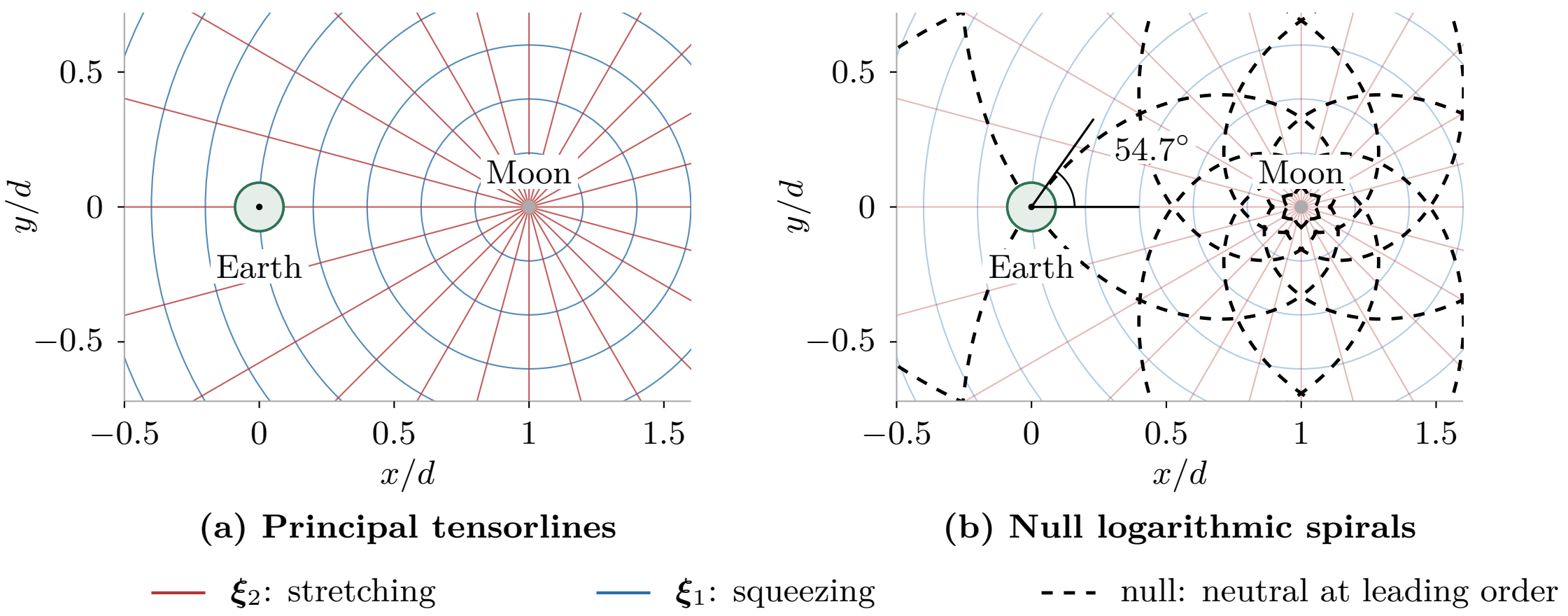
\begin{figure}[t!]
    \centering
    \begin{tikzpicture}[x=2.55cm,y=2.55cm,font=\footnotesize]
        \foreach \panel in {0,1}{
            \pgfmathsetlengthmacro{\panelshift}{\panel*7.3cm}
            \begin{scope}[xshift=\panelshift]
                \ifnum\panel=0\relax
                    \def\fieldopacity{0.8}
                \else
                    \def\fieldopacity{0.35}
                \fi
                \begin{scope}
                    \clip (-.5,-.72) rectangle (1.6,.72);
                    \foreach \rad in {.2,.4,...,2.4}{
                        \draw[squeeze,opacity=\fieldopacity,line width=.4pt]
                            (1,0) circle (\rad);
                    }
                    \foreach \angle in {0,15,...,165}{
                        \draw[stretch,opacity=\fieldopacity,line width=.4pt]
                            ({1-3*cos(\angle)},{-3*sin(\angle)})
                            --({1+3*cos(\angle)},{3*sin(\angle)});
                    }
                    \ifnum\panel=1\relax
                        \foreach \sense in {-1,1}{
                            \foreach \phase in {0,60,...,300}{
                                \draw[black,dashed,line width=.9pt]
                                    plot[domain=-2.9:1.03,samples=220,variable=\u]
                                    ({1+exp(\u)*cos(\phase+\sense*sqrt(2)*\u*180/pi)},
                                     {exp(\u)*sin(\phase+\sense*sqrt(2)*\u*180/pi)});
                            }
                        }
                    \fi
                    \filldraw[fill=earth!12,draw=earth,line width=.7pt]
                        (0,0) circle (.09);
                    \fill (0,0) circle (.012);
                    \fill[gray!65] (1,0) circle (.025);
                \end{scope}
                \draw[gray!65] (-.5,-.72)--(1.6,-.72);
                \draw[gray!65] (-.5,-.72)--(-.5,.72);
                \foreach \x/\lab in {-.5/{-0.5},0/0,.5/0.5,1/1,1.5/1.5}{
                    \draw (\x,-.72)--(\x,-.745);
                    \node[below] at (\x,-.745) {$\lab$};
                }
                \foreach \y/\lab in {-.5/{-0.5},0/0,.5/0.5}{
                    \draw (-.5,\y)--(-.525,\y);
                    \node[left] at (-.525,\y) {$\lab$};
                }
                \node[below] at (.55,-.88) {$x/d$};
                \node[rotate=90] at (-.88,0) {$y/d$};
                \node[fill=white,inner sep=1.5pt] at (0,-.22) {Earth};
                \node[fill=white,inner sep=1.5pt] at (1,.13) {Moon};
                \ifnum\panel=1\relax
                    \draw[black,line width=.6pt] (0,0)--(.4,0);
                    \draw[black,line width=.6pt]
                        (0,0)--({.4/sqrt(3)},{.4*sqrt(2/3)});
                    \draw[black,line width=.5pt]
                        (.16,0) arc[start angle=0,end angle=54.7356,radius=.16];
                    \node[anchor=west,fill=white,inner sep=1pt] at (.29,.22)
                        {$54.7^\circ$};
                \fi
                \ifnum\panel=0\relax
                    \node[anchor=north,align=center] at (.55,-1.08)
                        {\textbf{(a) Principal tensorlines}};
                \else
                    \node[anchor=north,align=center] at (.55,-1.08)
                        {\textbf{(b) Null logarithmic spirals}};
                \fi
            \end{scope}
        }
        \draw[stretch,line width=.8pt] (-.4,-1.43)--(-.2,-1.43);
        \node[anchor=west] at (-.16,-1.43) {$\xxi_2$: stretching};
        \draw[squeeze,line width=.8pt] (1.25,-1.43)--(1.45,-1.43);
        \node[anchor=west] at (1.49,-1.43) {$\xxi_1$: squeezing};
        \draw[black,dashed,line width=.9pt] (2.9,-1.43)--(3.1,-1.43);
        \node[anchor=west] at (3.14,-1.43) {null: neutral at leading order};
    \end{tikzpicture}
    \caption{Principal and null geometry of the planar lunar tidal tensor. (a) Radial stretchlines (red) and circular squeezelines (blue), centered on the Moon at $(d,0)$, with Earth at $(0,0)$. (b) The same field with the two families of null logarithmic spirals (thick black dashed curves for both senses). Their tangents make the constant acute angle $\arctan\sqrt{2}\approx54.7^\circ$ with radial lines; the angle is marked at Earth. For particles released at rest, tangent line elements have no quadratic-in-time length change. These curves are not particle trajectories, sea-surface contours, or boundaries of the tidal bulges. Earth's outline is enlarged; line spacing does not encode deformation strength.}
    \label{fig:field}
\end{figure}

Borrowing LCS terminology, call integral curves of $\xxi_2$ \emph{stretchlines} and those of $\xxi_1$ \emph{squeezelines}. They satisfy
\begin{equation}
    \frac{\dd\boldsymbol\gamma_i}{\dd\ell}
    =\xxi_i(\boldsymbol\gamma_i(\ell)),\qquad i=1,2,
    \label{eq:tensorline}
\end{equation}
with $\ell$ a spatial arclength parameter, not time. Eigenvector signs only reverse traversal; the tensorlines themselves are unoriented. In polar coordinates centered on the Moon, our representatives are $\xxi_1=-\mathbf e_\theta$ and $\xxi_2=-\mathbf e_\rho$. Equation~\eqref{eq:tensorline} therefore integrates to circles $\rho=\rho_0$ for squeezelines and rays $\theta=\theta_0$ for stretchlines. The Moon's singular location is excluded. The two families form the orthogonal field shown in Fig.~\ref{fig:field}(a). Panel (b) adds the null curves of the associated Lorentzian metric, developed below in Sec.~\ref{sec:nullgeometry}.

Define
\begin{equation}
    k:=GM/d^3.
    \label{eq:defk}
\end{equation}
At Earth's center, the tensor and its eigenvectors reduce to
\begin{equation}
    \TT(\mathbf0)=k\begin{pmatrix}2&0\\0&-1\end{pmatrix}, \qquad \xxi_1(\mathbf0)=\begin{pmatrix}0\\1\end{pmatrix}, \qquad \xxi_2(\mathbf0)=\begin{pmatrix}1\\0\end{pmatrix}.
    \label{eq:origin}
\end{equation}

For an infinitesimal separation $\delta\rr=(\delta x,\delta y)$ from the reference particle at Earth's center, Eq.~\eqref{eq:variation} therefore gives two independent components:
\begin{itemize}
    \item Along the Earth--Moon axis, $\delta\ddot x=2k\,\delta x$. The relative acceleration has the same sign as the separation. On the near side, the particle accelerates toward the Moon more strongly than the center; on the far side, it accelerates less strongly. With zero initial relative velocity, the magnitude of the longitudinal separation initially increases on either side. These are the two outward ends of one stretching deformation.

    \item Perpendicular to the Earth--Moon axis, $\delta\ddot y=-k\,\delta y$. The relative acceleration opposes the separation, pointing toward the axis from above and below. With zero initial relative velocity, the magnitude of the transverse separation initially decreases, giving squeezing.
\end{itemize}

These component equations describe infinitesimal separations about the center. Since the lunar tidal tensor varies on the distance scale $d$, its Taylor expansion gives
\begin{equation}
    \TT(\rr)=\TT(\mathbf0)+O\!\left(k\,\frac{|\rr|}{d}\right), \qquad \frac{|\rr|}{d}\ll1.
    \label{eq:constant-tensor}
\end{equation}
Thus, for $|\rr|/d\ll1$, neglecting the correction in Eq.~\eqref{eq:constant-tensor} gives $\TT(\rr)\approx\TT(\mathbf0)$, extending the same stretching and squeezing directions throughout Earth's neighborhood to leading order. In this constant-tensor approximation, the stretchlines are the parallel horizontal lines $y=\mathrm{constant}$, and the squeezelines are the parallel vertical lines $x=\mathrm{constant}$. The exact lunar tensorlines are instead Moon-centered rays and circles, which appear nearly horizontal and vertical on Earth's scale.

\section{Short-time deformation}

The acceleration in Eq.~\eqref{eq:relative} is autonomous. Write it as $\mathbf a(\rr)=\gg(\rr)-\gg(\mathbf0)$ and choose release at $t=0$. Its full flow acts on position and velocity: $\dot\rr=\mathbf v$, $\dot{\mathbf v}=\mathbf a(\rr)$. To define a map from initial positions alone, we must also prescribe the initial velocities. We take $\rr(0)=\rr_0$ and $\dot\rr(0)=\mathbf0$ for every particle and define $F_t(\rr_0):=\rr(t;\rr_0,\mathbf0)$. This map takes initial positions to positions at time $t$. Unlike the full phase-space flow, it generally does not satisfy $F_{t+s}=F_t\circ F_s$: applying $F_t$ to the position reached at time $s$ would prescribe zero velocity there, whereas the continuing trajectory generally has nonzero velocity. Away from the Moon, the acceleration field is smooth, so the phase-space trajectories depend smoothly on time and initial conditions. Consequently, $DF_t(\rr_0)$ is continuous and equals $\Id$ at $t=0$. Its determinant therefore remains nonzero for sufficiently small $t$, and the inverse function theorem makes $F_t$ locally invertible near each initial position.

Integrating twice and using $\rr(0)=\rr_0$ and $\dot\rr(0)=\mathbf0$ yields
\begin{equation}
    F_t(\rr_0)
    =\rr_0+\int_0^t\int_0^{t''}
        \mathbf a(F_{t'}(\rr_0))\,\dd t'\,\dd t''
    =\rr_0+\int_0^t(t-t')
        \mathbf a(F_{t'}(\rr_0))\,\dd t'.
    \label{eq:integralmap}
\end{equation}
The second equality follows by reversing the integration order over $0\leq t'\leq t''\leq t$ and integrating with respect to $t''$. This exact integral equation provides the short-time expansion without solving the nonlinear trajectories. 

For a small forward time $\tau>0$, the zero initial velocity implies $F_\tau(\rr_0)-\rr_0=O(\tau^2)$ and hence $\mathbf a(F_\tau(\rr_0))=\mathbf a(\rr_0)+O(\tau^2)$. Substituting the corresponding estimate for the integrand into Eq.~\eqref{eq:integralmap}, evaluated at $t=\tau$, gives
\begin{equation}
    F_\tau(\rr_0)
    =\rr_0+\tfrac12\tau^2\mathbf a(\rr_0)+O(\tau^4).
    \label{eq:positionexpansion}
\end{equation}
Differentiating with respect to the initial position yields
\begin{equation}
    D F_\tau(\rr_0)
    =\Id+\tfrac12\tau^2\TT(\rr_0)+O(\tau^4),
    \label{eq:mapexpansion}
\end{equation}
since $D\mathbf a(\rr)=D\gg(\rr)=\TT(\rr)$.

The right Cauchy--Green tensor of this configuration map is
\begin{equation}
    \CC_\tau(\rr_0)
    =D F_\tau(\rr_0)^\transpose D F_\tau(\rr_0)
    =\Id+\tau^2\TT(\rr_0)+O(\tau^4).
    \label{eq:shorttime}
\end{equation}
Indeed, the two quadratic contributions in the matrix product sum to $\tfrac12\tau^2(\TT^\transpose+\TT)=\tau^2\TT$, while their product is $O(\tau^4)$. This establishes the connection between the tidal tensor and leading-order material deformation.

As in LCS theory, order the eigenpairs of $\CC_\tau$ as $0<\lambda_1\leq\lambda_2$. An initial line element aligned with $\xxi_i$ changes length by the factor $\sqrt{\lambda_i}$ over $[0,\tau]$. Its $\xxi_2$ tensorlines are stretchlines and its $\xxi_1$ tensorlines are squeezelines \citep{HallerBeronVera2012,FarazmandHaller2013,FarazmandEtAl2014,BeronVera2015}. Tangential stretching and normal stability must be distinguished: a stretchline attracts normally when $\lambda_1<1$, whereas a squeezeline repels normally when $\lambda_2>1$. Both inequalities hold here for sufficiently small positive $\tau$, wherever the tidal eigenvalues have opposite signs. These tensorlines therefore describe normally attracting and repelling material curves, respectively. Their variational selection as hyperbolic LCSs involves proximity to least-stretching geodesics \citep[Sec.~6.1]{HallerBeronVera2012} or stationary averaged Lagrangian shear with fixed endpoints, together with extremal averaged stretching among neighboring tensorlines \citep[Sec.~7.1]{FarazmandEtAl2014}.

In continuum mechanics, the departure of $\CC_\tau$ from the identity defines the Green--Lagrange strain tensor $\mathsf E_\tau=\tfrac12(\CC_\tau-\Id)$ \citep{TruesdellNoll2004}. Equation~\eqref{eq:shorttime} gives $\mathsf E_\tau(\rr_0)=\tfrac12\tau^2\TT(\rr_0)+O(\tau^4)$: the tidal tensor is twice the coefficient of the quadratic term in the short-time strain expansion. Wherever the tidal eigenvalues are distinct, the principal directions of $\CC_\tau$ converge to those of $\TT$ as $\tau\downarrow0$. In increasing order, their eigenvalues satisfy $\lambda_i^{\CC}=1+\tau^2\lambda_i^{\TT}+O(\tau^4)$, with the tidal eigenvalues evaluated at $\rr_0$. The radial and circular tensorlines in Fig.~\ref{fig:field}(a) therefore give the limiting principal stretching and squeezing directions of the released particle family.

\begin{remark}
    The present short-time limit differs from the first-order limit underlying hyperbolic objective Eulerian coherent structures (OECSs) \textup{\citep{SerraHaller2016}}. For a smooth velocity field $\vv(\rr,t)$, the Cauchy--Green expansion begins with $\Id+2\tau\mathsf S(\rr_0,0)+O(\tau^2)$, where $\mathsf S=\tfrac12(D\vv+(D\vv)^\transpose)$ is the rate-of-strain tensor. Our released family induces a generally time-dependent velocity field even though $\mathbf a(\rr)$ is autonomous. At release, $\vv(\rr,0)=\mathbf0$ throughout the family, so $\mathsf S(\rr,0)=\mathbf0$: its eigenvalues coincide and its principal directions are undetermined. The quadratic term in Eq.~\eqref{eq:shorttime} resolves this degeneracy wherever the tidal eigenvalues are distinct, determining limiting stretching and squeezing directions that the first-order hyperbolic OECS construction does not distinguish at release.
\end{remark}

\subsection{Null directions and logarithmic spirals}
\label{sec:nullgeometry}

The short-time relation also defines a Lorentzian geometry on the configuration plane. Since the symmetric tensor $\TT(\rr)$ has eigenvalues $\lambda_1<0<\lambda_2$ at every finite distance from the Moon, the bilinear form
\begin{equation}
    \mathcal T_{\rr}(\tt,\tt')
    :=\tt^\transpose\TT(\rr)\tt'
    \label{eq:tidalmetric}
\end{equation}
is nondegenerate with signature $(-,+)$. This metric acts on spatial separation vectors, not spacetime intervals. For an infinitesimal material segment with initial separation $\delta\rr(0)=\ell_0\tt$, where $|\tt|=1$ and $\ell_0$ is its initial length, Eq.~\eqref{eq:shorttime} gives
\begin{equation}
    \frac{|\delta\rr(\tau)|^2-\ell_0^2}{\ell_0^2}
    =\tau^2\mathcal T_{\rr_0}(\tt,\tt)+O(\tau^4).
    \label{eq:neutralstretching}
\end{equation}
The sign of $\mathcal T_{\rr_0}(\tt,\tt)$ distinguishes leading-order lengthening from shortening. In a null direction, stretching along the Moon-centered radial direction and squeezing perpendicular to it cancel in the quadratic change of length. With zero initial relative velocity, the separation evolves as
\begin{equation}
    \delta\rr(\tau)
    =\ell_0\left(\tt+\tfrac12\tau^2\TT(\rr_0)\tt+O(\tau^4)\right).
\end{equation}
Nullity, $\tt^\transpose\TT(\rr_0)\tt=0$, makes the nonzero quadratic correction perpendicular to the initial segment. Its orientation therefore changes at quadratic order in time, while its length has no quadratic change.

In Moon-centered polar coordinates, $x=d+\rho\cos\theta$ and $y=\rho\sin\theta$, the metric takes the form
\begin{equation}
    \mathcal T=\frac{GM}{\rho^3}
    \left(2\,\dd\rho^2-\rho^2\,\dd\theta^2\right).
    \label{eq:polarmetric}
\end{equation}
At each point, the null cone consists of two lines in the tangent plane, specified by $\rho\,\dd\theta=\pm\sqrt{2}\,\dd\rho$. Integrating these direction fields gives
\begin{equation}
    \theta-\theta_0
    =\pm\sqrt{2}\log\!\frac{\rho}{\rho_0}.
    \label{eq:nullspirals}
\end{equation}
These are the two families of logarithmic spirals in Fig.~\ref{fig:field}(b): their tangents follow the local null directions. They meet radial stretchlines at the constant acute angle $\alpha=\arctan|\rho\,\dd\theta/\dd\rho|=\arctan\sqrt{2} \approx 54.7^\circ$ and circular squeezelines at the complementary angle $\pi/2-\alpha=\arctan(1/\sqrt{2}) \approx 35.3^\circ$. Their shape is independent of the positive factor $GM$. 

\begin{remark}
    In two-dimensional Lorentzian geometry, a curve following a null direction has covariant acceleration parallel to its tangent. A suitable reparametrization makes this acceleration vanish, giving the geodesic equation \textup{\citep[Ch.~3]{ONeill1983}}.
\end{remark}

A fixed smooth compact spiral segment, away from the Moon and populated with particles released at rest, has no quadratic change in material length, although its shape can change and quartic stretching generally remains. These spirals describe neither particle trajectories nor contours of ocean height. They separate lengthening and shortening orientations without altering the two-bulge interpretation, which follows from the two ends of longitudinal stretching; the Earth-scale consequences are illustrated in Sec.~\ref{sec:earthdeformation}. Since the spirals do not close, this tidal metric admits no closed null-geodesic loops.

To distinguish the finite-time geometry from its limit, define
\begin{equation}
    \TT_\tau(\rr_0):=\frac{\CC_\tau(\rr_0)-\Id}{\tau^2}
    =\TT(\rr_0)+O(\tau^2),
    \qquad
    \mathcal T_{\tau,\rr_0}(\tt,\tt')
    :=\tt^\transpose\TT_\tau(\rr_0)\tt'.
    \label{eq:finitetimemetric}
\end{equation}
On compact regions away from the Moon, $\mathcal T_\tau$ has Lorentzian signature for sufficiently small $\tau>0$. Its null directions satisfy $\tt^\transpose\CC_\tau(\rr_0)\tt=|\tt|^2$, expressing exact preservation of infinitesimal line-element length over $[0,\tau]$. As $\tau\downarrow0$, $\mathcal T_\tau$ converges to $\mathcal T$. The logarithmic spirals thus describe the limiting null curves of this unit-stretch geometry, rather than exact finite-time length-preserving curves in general.

\subsection{Earth-scale deformation and zero radial displacement}
\label{sec:earthdeformation}

Let $r_\odot$ denote Earth's radius. Since $r_\odot\ll d$, Eq.~\eqref{eq:constant-tensor} justifies replacing the tidal tensor throughout Earth's neighborhood by $\TT(\mathbf0)$, given in Eq.~\eqref{eq:origin}, and the relative acceleration in Eq.~\eqref{eq:relative} by $\TT(\mathbf0)\rr$ to leading order. This spatial approximation is separate from the short-time expansion. In the constant-tensor approximation, the null spirals reduce to two families of parallel straight lines, $y=\pm\sqrt2\,x+b$, inclined at approximately $\pm54.7^\circ$ to the Earth--Moon axis. The two lines through Earth's center have $b=0$ and intersect Earth's circumference at B, D, F, and H in Fig.~\ref{fig:earthdeformation}. These four locations connect three distinct effects: deformation of material disks, displacement of their centers, and zero equilibrium elevation.

\subsubsection{Deformation of material disks}

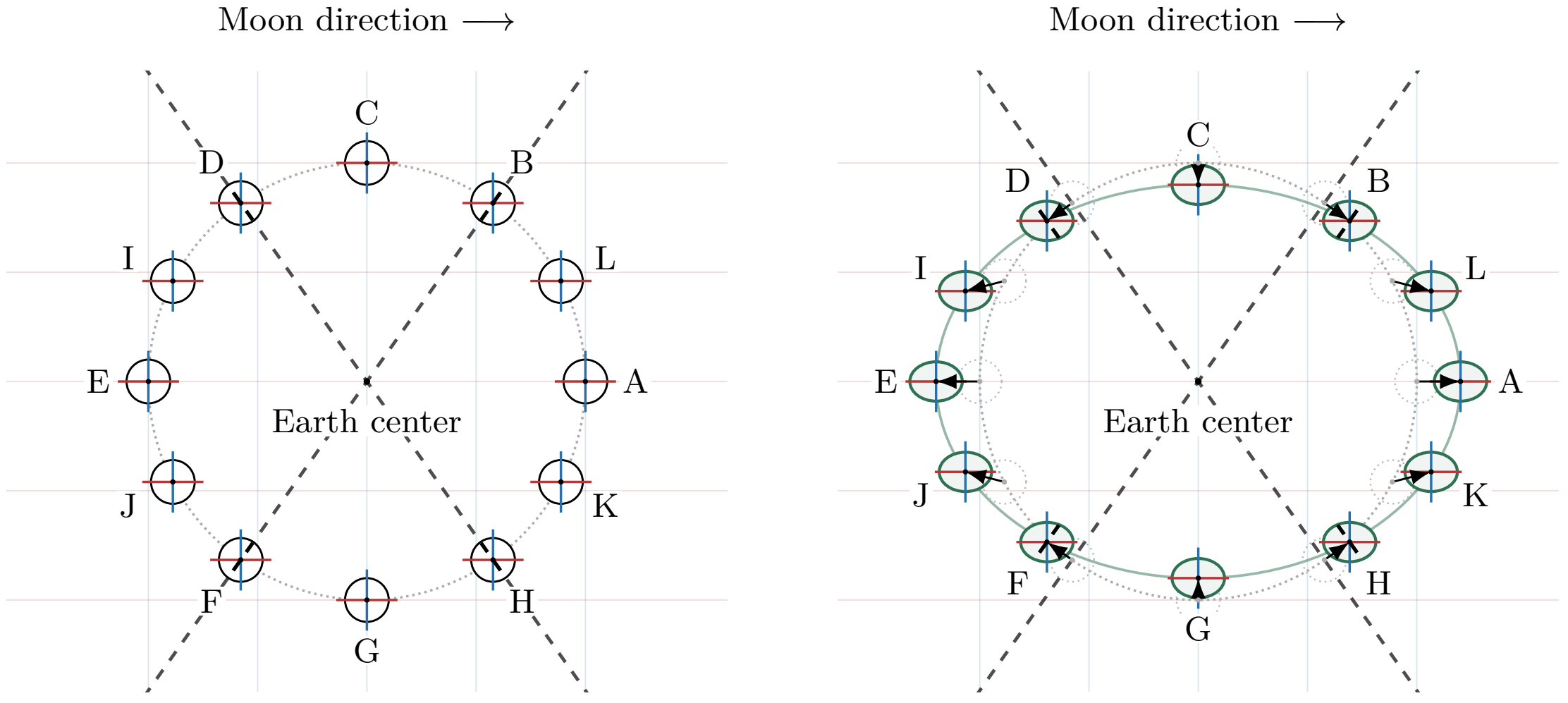
\begin{figure}[htbp]
    \centering
    \begin{tikzpicture}[x=2.05cm,y=2.05cm,>=Latex,font=\footnotesize]
        \def\eps{0.10}
        \pgfmathsetmacro{\alphaNull}{acos(1/sqrt(3))}
        \foreach \panel in {0,1}{
            \pgfmathsetlengthmacro{\panelshift}{\panel*7.8cm}
            \begin{scope}[xshift=\panelshift]
                \pgfmathsetmacro{\ax}{1+\panel*2*\eps}
                \pgfmathsetmacro{\ay}{1-\panel*\eps}
                \begin{scope}
                    \clip (-1.65,-1.42) rectangle (1.65,1.42);
                    \foreach \level in {-1,-.5,0,.5,1}{
                        \draw[stretch,opacity=.18,line width=.35pt]
                            (-1.65,\level)--(1.65,\level);
                        \draw[squeeze,opacity=.18,line width=.35pt]
                            (\level,-1.42)--(\level,1.42);
                    }
                    \foreach \sense in {-1,1}{
                        \draw[black,dashed,line width=.9pt,opacity=.7]
                            (-1.2,{-\sense*sqrt(2)*1.2})
                            --(1.2,{\sense*sqrt(2)*1.2});
                    }
                \end{scope}
                \draw[gray!65,densely dotted,line width=.65pt]
                    (0,0) circle (1);
                \ifnum\panel=1\relax
                    \draw[earth,opacity=.5,line width=.7pt]
                        (0,0) ellipse [x radius=\ax,y radius=\ay];
                \fi
                \foreach \angle/\name/\isnull in {
                    0/A/0,\alphaNull/B/1,90/C/0,
                    {180-\alphaNull}/D/1,180/E/0,
                    {180+\alphaNull}/F/1,270/G/0,
                    {360-\alphaNull}/H/1,
                    {180-\alphaNull/2}/I/0,
                    {180+\alphaNull/2}/J/0,
                    {360-\alphaNull/2}/K/0,
                    {\alphaNull/2}/L/0}{
                    \pgfmathsetmacro{\px}{cos(\angle)}
                    \pgfmathsetmacro{\py}{sin(\angle)}
                    \pgfmathsetmacro{\qx}{\ax*\px}
                    \pgfmathsetmacro{\qy}{\ay*\py}
                    \ifnum\panel=1\relax
                        \draw[gray!55,densely dotted,line width=.45pt]
                            (\px,\py) circle (.10);
                    \fi
                    \ifnum\panel=0\relax
                        \draw[black,fill=white,line width=.55pt]
                            (\qx,\qy) circle (.10);
                    \else
                        \draw[earth,fill=earth!7,line width=.8pt]
                            (\qx,\qy) ellipse
                            [x radius={.10*\ax},y radius={.10*\ay}];
                    \fi
                    \draw[stretch,line width=.65pt]
                        ({\qx-.14},\qy)--({\qx+.14},\qy);
                    \draw[squeeze,line width=.65pt]
                        (\qx,{\qy-.14})--(\qx,{\qy+.14});
                    \ifnum\isnull=1\relax
                        \draw[black,dashed,line width=1pt]
                            ({\qx-.095*\px},{\qy-.095*\py})
                            --({\qx+.095*\px},{\qy+.095*\py});
                    \fi
                    \ifnum\panel=1\relax
                        \draw[->,black,line width=.55pt]
                            (\px,\py)--(\qx,\qy);
                        \fill[gray!65] (\px,\py) circle (.012);
                    \fi
                    \fill (\qx,\qy) circle (.012);
                    \node[fill=white,inner sep=.7pt]
                        at ({\qx+.23*\px},{\qy+.23*\py}) {\name};
                }
                \fill (0,0) circle (.015);
                \node[fill=white,inner sep=1pt] at (0,-.18) {Earth center};
                \node at (0,1.66) {Moon direction $\longrightarrow$};
                \ifnum\panel=0\relax
                    \node[anchor=north,align=center] at (0,-1.52)
                        {\textbf{(a) Initial material disks}};
                \else
                    \node[anchor=north,align=center] at (0,-1.52)
                        {\textbf{(b) Leading short-time deformation}};
                \fi
            \end{scope}
        }
    \end{tikzpicture}
    \caption{Earth-scale deformation in the constant-tensor approximation. (a) Twelve circular material disks initially have centers on Earth's circumference. Red and blue segments show horizontal stretching and vertical squeezing directions; thick black dashed lines show null directions. (b) Equation~\eqref{eq:earthmap}, illustrated at $\tfrac12k\tau^2=0.10$, takes all disks to ellipses and the initial circumference to an ellipse (green). Gray dotted outlines mark initial configurations; black arrows join initial and displaced centers. At B, D, F, and H, the leading center displacement is tangential to the initial circumference. Direction segments in both panels sample the same fixed fields and are not advected with the material. I, J, K, and L initially bisect the arcs D--E, E--F, H--A, and A--B, respectively. Coordinates are scaled by $r_\odot$; disk sizes and deformation are exaggerated. The figure describes leading short-time deformation, not sea-surface elevation.}
    \label{fig:earthdeformation}
\end{figure}

In the constant-tensor approximation, release with zero initial relative velocity gives
\begin{equation}
    F_\tau(\rr_0)
    =\rr_0+\tfrac12\tau^2\TT(\mathbf0)\rr_0
    +O(k^2\tau^4|\rr_0|).
    \label{eq:earthmap}
\end{equation}
For the Cauchy--Green tensor of this constant-tensor model, the principal length factors are
\begin{equation}
    \sqrt{\lambda_1^{\CC}}
    =1-\tfrac12 k\tau^2+O(k^2\tau^4),
    \qquad
    \sqrt{\lambda_2^{\CC}}
    =1+k\tau^2+O(k^2\tau^4).
\end{equation}
Thus an initially circular material disk of radius $\ell_0$ becomes an ellipse whose vertical and horizontal semiaxes are obtained by multiplying $\ell_0$ by the first and second length factors above, respectively. These factors describe squeezing and stretching relative to the moving disk center and are independent of its initial position in this approximation. Figure~\ref{fig:earthdeformation} illustrates this deformation for twelve disks initially centered on Earth's circumference. All have the same deformation relative to their moving centers, including those at B, D, F, and H.

A null orientation does not prevent this deformation. An infinitesimal material diameter initially aligned with a null direction turns at quadratic order without a quadratic change in length, as shown in Sec.~\ref{sec:nullgeometry}. Other diameters lengthen or shorten, so the disk becomes an ellipse. The dashed segments in the figure show the fixed null directions of $\TT(\mathbf0)$, not advected material diameters. They describe vanishing quadratic extension; exact finite-time unit-stretch directions instead belong to $\mathcal T_\tau$.

\subsubsection{Displacement of disk centers}

To describe the motion of any disk center, write $\rr_0=r_\odot\mathbf n$, where $\mathbf n=(\cos\psi,\sin\psi)$ and $\psi$ is measured from the Earth--Moon axis. Here radial means outward from Earth's center, along $\mathbf n$; tangential means perpendicular to $\mathbf n$, along Earth's circumference. These directions differ from the Moon-centered radial and tangential directions of Sec.~\ref{sec:nullgeometry}.

Equation~\eqref{eq:earthmap} gives the center displacement
\begin{equation}
    F_\tau(r_\odot\mathbf n)-r_\odot\mathbf n
    =\tfrac12kr_\odot\tau^2
        \begin{pmatrix}2\cos\psi\\-\sin\psi\end{pmatrix}
    +O(r_\odot k^2\tau^4).
\end{equation}
At A and E it points horizontally outward, while at C and G it points vertically inward. Elsewhere it generally has both Earth-centered radial and tangential components. The radial component of the initial relative acceleration is
\begin{equation}
    \mathbf n^\transpose\TT(\mathbf0)\rr_0
    =kr_\odot(3\cos^2\psi-1),
    \label{eq:radialacceleration}
\end{equation}
and the corresponding change in distance from Earth's center is
\begin{equation}
    |F_\tau(r_\odot\mathbf n)|-r_\odot
    =\tfrac12kr_\odot\tau^2(3\cos^2\psi-1)
    +O(r_\odot k^2\tau^4).
    \label{eq:nullradialdisplacement}
\end{equation}
The leading radial displacement is outward on the arcs containing A and E and inward on those containing C and G. It vanishes at $\psi=\alpha,\pi-\alpha,\pi+\alpha,2\pi-\alpha$, with $\alpha=\arctan\sqrt2\approx54.7^\circ$, corresponding to B, D, F, and H.

At these four locations, the Earth-centered radial direction is null: $\mathbf n^\transpose\TT(\mathbf0)\mathbf n=0$. The initial relative acceleration remains nonzero and is tangential, producing a leading tangential displacement. These centers therefore have no quadratic radial displacement, but they are not fixed; quartic terms generally move them off the initial circumference.

\begin{remark}
    In the constant-tensor model, the exact released solution is $x(t)=x_0\cosh(\sqrt{2k}\,t)$ and $y(t)=y_0\cos(\sqrt{k}\,t)$. Centers initially on either coordinate axis therefore remain on that axis beyond the short-time approximation. The four locations with null radial directions have no corresponding invariance. For the exact lunar field, only the Earth--Moon axis remains invariant.
\end{remark}

\subsubsection{Connection with equilibrium elevation}

Zero tidal elevation requires a separate equilibrium model. The lunar acceleration derives from the positive potential $GM/|\rr-\RR|$. Subtracting its constant and linear Taylor terms at Earth's center fixes the potential's reference value and removes the common acceleration $\gg(\mathbf0)$. The remaining quadratic tide-generating potential is $U(\rr)=\tfrac12\rr^\transpose\TT(\mathbf0)\rr$, with $\nabla U=\TT(\mathbf0)\rr$.

For a spherical Earth covered by an equilibrium ocean, the surface is an equipotential. Its leading elevation satisfies $g_\odot\eta_{\mathrm{eq}}=U$, where $g_\odot$ is Earth's surface gravitational acceleration \citep{Hendershott2005}. Hence
\begin{equation}
    \eta_{\mathrm{eq}}(\psi)
    =\frac{kr_\odot^2}{2g_\odot}(3\cos^2\psi-1).
    \label{eq:equilibriumzeros}
\end{equation}
The same angular factor appears in Eqs.~\eqref{eq:radialacceleration}--\eqref{eq:equilibriumzeros}. Consequently, B, D, F, and H have both zero leading radial displacement in the freely falling model and zero elevation in the equilibrium model. On the full sphere, these locations belong to two circles around the Earth--Moon axis. Their tangential acceleration remains nonzero, and their surrounding material still deforms.

This coincidence follows from the shared quadratic potential, not from an equivalence of the two physical models. It also relies on the spatial approximation used here: in the exact lunar field, the relative acceleration and the local tidal tensor must be distinguished. Two null directions exist at every point, but the Earth-centered radial direction aligns with one of them only at these four locations on the circumference.

\section{Concluding remarks}

Tendex lines and the stretchlines and squeezelines of finite-time deformation share a tensorline construction. For the autonomous lunar model and particles released with zero initial velocity relative to a freely falling reference, their dynamical connection is expressed by $\CC_\tau(\rr_0)=\Id+\tau^2\TT(\rr_0)+O(\tau^4)$. The tidal tensor determines both the leading Green--Lagrange strain and the limiting principal deformation directions. Near Earth, this geometry reveals the two outward ends of longitudinal stretching, accompanied by transverse squeezing. The same tensor defines the Lorentzian metric $\mathcal T$, whose null directions integrate to logarithmic spirals. Material segments initially aligned with these directions turn at quadratic order without a quadratic change in length, because their relative acceleration is perpendicular to their separation.

The Earth-scale disk construction separates this local deformation from the displacement of a disk's center. In the constant-tensor approximation, four locations on Earth's circumference have null Earth-centered radial directions. Their centers have tangential leading displacements, while the surrounding disks deform into ellipses. These locations also have zero degree-two equilibrium elevation: radial acceleration and tide-generating potential contain the same quadratic form. This coincidence connects the freely falling and equilibrium models without equating their physical responses. In particular, zero equilibrium elevation implies neither a stationary particle nor an undeformed material region. Predicting the ocean's dynamical response requires additional physics.

The initial velocity prescription is essential to the deformation interpretation. Free fall permits arbitrary initial velocity and does not imply motion along gravitational acceleration. Zero initial relative velocity makes the leading relative displacement quadratic in time and aligned with the initial relative acceleration. A spatially uniform nonzero initial relative velocity preserves the same quadratic deformation tensor but generally introduces a cubic correction; spatial variations in initial velocity can introduce a linear contribution through the rate-of-strain tensor. Even for release at relative rest, the limiting correspondence does not imply equality of the tidal and Cauchy--Green tensorline fields over finite intervals.

The bilinear form represented by $\CC_\tau-\Id$ also appears, up to a constant factor, as the unit-stretch member of the Lorentzian metric family used to identify coherent Lagrangian vortex boundaries \citep{HallerBeronVera2013,HallerBeronVera2014}, motivating their analogy with black holes. Dividing this matrix by $\tau^2$ gives the tidal tensor in the short-time limit, identifying $\mathcal T$ as the leading counterpart of that finite-time geometry. Its null curves can be obtained either by integrating the null direction fields or by solving the metric's null condition directly. The explicit lunar metric makes this construction analytically tractable, yielding logarithmic spirals. Their open geometry provides no closed null-geodesic vortex boundary, but the common construction establishes a further dynamical connection between gravitational tidal geometry and Lagrangian deformation.

\section*{Acknowledgments}

I thank the Intercultural Outreach Initiative (IOI) in Isabela, Gal\'apagos, for their hospitality during the writing of this work. The study was motivated by my reading of Kip Thorne's book on the movie \emph{Interstellar} and a desire to clarify the two-bulge tidal deformation, whose explanation I have long found elusive.

\section*{Funding}

This work was carried out without any specific grant from funding agencies.

\section*{Author Declarations}

\subsection*{Conflict of Interest}

The author has no conflict of interest to disclose.

\subsection*{Author Contributions}

F.J.B.-V.: Conceptualization, Formal analysis, Investigation, Visualization, Writing -- original draft, Writing -- review \& editing.

\section*{Data and Software Availability}

No data or software were generated or used in this study.

\bibliographystyle{plainnat}
\bibliography{tides}

@book{ONeill1983,
    author    = {O'Neill, Barrett},
    title     = {Semi-Riemannian Geometry with Applications to Relativity},
    series    = {Pure and Applied Mathematics},
    volume    = {103},
    publisher = {Academic Press},
    address   = {New York},
    year      = {1983}
}

@book{TruesdellNoll2004,
    author    = {Truesdell, Clifford and Noll, Walter},
    title     = {The Non-Linear Field Theories of Mechanics},
    edition   = {3},
    year      = {2004},
    publisher = {Springer},
    address   = {Berlin, Heidelberg},
    doi       = {10.1007/978-3-662-10388-3}
}

@incollection{BeronVera2015,
    author    = {Beron-Vera, Francisco J.},
    title     = {Flow Coherence: Distinguishing Cause from Effect},
    booktitle = {Selected Topics of Computational and Experimental Fluid Mechanics},
    editor    = {Klapp, Jaime and Ru{\'i}z Chavarr{\'i}a, Gerardo and Medina Ovando, Abraham and L{\'o}pez Villa, Abel and Sigalotti, Leonardo Di G.},
    series    = {Environmental Science and Engineering},
    publisher = {Springer},
    address   = {Cham},
    year      = {2015},
    pages     = {81--89},
    doi       = {10.1007/978-3-319-11487-3_4}
}

@book{CreightonAnderson2011,
    author    = {Creighton, Jolien D. E. and Anderson, Warren G.},
    title     = {Gravitational-Wave Physics and Astronomy: An Introduction to Theory, Experiment and Data Analysis},
    publisher = {Wiley-VCH},
    address   = {Weinheim},
    year      = {2011}
}

@article{FarazmandEtAl2014,
    author  = {Farazmand, Mohammad and Blazevski, Daniel and Haller, George},
    title   = {Shearless transport barriers in unsteady two-dimensional flows and maps},
    journal = {Physica D},
    volume  = {278--279},
    pages   = {44--57},
    year    = {2014},
    doi     = {10.1016/j.physd.2014.03.008}
}

@article{FarazmandHaller2013,
    author  = {Farazmand, Mohammad and Haller, George},
    title   = {Attracting and repelling {Lagrangian} coherent structures from a single computation},
    journal = {Chaos},
    volume  = {23},
    pages   = {023101},
    year    = {2013},
    doi     = {10.1063/1.4800210}
}

@book{Haller2023,
    author    = {Haller, George},
    title     = {Transport Barriers and Coherent Structures in Flow Data: Advective, Diffusive, Stochastic and Active Methods},
    publisher = {Cambridge University Press},
    year      = {2023},
    doi       = {10.1017/9781009225199}
}

@article{HallerBeronVera2012,
    author  = {Haller, George and Beron-Vera, Francisco J.},
    title   = {Geodesic theory of transport barriers in two-dimensional flows},
    journal = {Physica D},
    volume  = {241},
    pages   = {1680--1702},
    year    = {2012},
    doi     = {10.1016/j.physd.2012.06.012}
}

@article{HallerBeronVera2013,
    author  = {Haller, George and Beron-Vera, Francisco J.},
    title   = {Coherent {Lagrangian} vortices: the black holes of turbulence},
    journal = {Journal of Fluid Mechanics},
    volume  = {731},
    pages   = {R4},
    year    = {2013},
    doi     = {10.1017/jfm.2013.391}
}

@article{HallerBeronVera2014,
    author  = {Haller, George and Beron-Vera, Francisco J.},
    title   = {Addendum to `Coherent {Lagrangian} vortices: the black holes of turbulence'},
    journal = {Journal of Fluid Mechanics},
    volume  = {755},
    pages   = {R3},
    year    = {2014},
    doi     = {10.1017/jfm.2014.441}
}

@incollection{Hendershott2005,
    author    = {Hendershott, Myrl},
    title     = {Introduction to ocean tides},
    booktitle = {2004 Program of Study: Tides},
    publisher = {Woods Hole Oceanographic Institution},
    year      = {2005},
    pages     = {1--19},
    note      = {Report WHOI-2005-08. Lecture notes prepared by Josefina Arraut and Anja Slim}
}

@book{Newton1999,
    author    = {Newton, Isaac},
    title     = {The {Principia}: Mathematical Principles of Natural Philosophy},
    publisher = {University of California Press},
    address   = {Berkeley},
    year      = {1999},
    note      = {Translated by I. Bernard Cohen and Anne Whitman, assisted by Julia Budenz. Originally published in 1687}
}

@article{NicholsEtAl2011,
    author  = {Nichols, David A. and Owen, Robert and Zhang, Fan and Zimmerman, Aaron and Brink, Jeandrew and Chen, Yanbei and Kaplan, Jeffrey D. and Lovelace, Geoffrey and Matthews, Keith D. and Scheel, Mark A. and Thorne, Kip S.},
    title   = {Visualizing spacetime curvature via frame-drag vortexes and tidal tendexes: General theory and weak-gravity applications},
    journal = {Physical Review D},
    volume  = {84},
    pages   = {124014},
    year    = {2011},
    doi     = {10.1103/PhysRevD.84.124014}
}

@article{SerraHaller2016,
    author  = {Serra, Mattia and Haller, George},
    title   = {Objective {Eulerian} coherent structures},
    journal = {Chaos},
    volume  = {26},
    pages   = {053110},
    year    = {2016},
    doi     = {10.1063/1.4951720}
}

@book{Thorne2014,
    author    = {Thorne, Kip S.},
    title     = {The Science of {Interstellar}},
    publisher = {W. W. Norton \& Company},
    address   = {New York},
    year      = {2014}
}

\end{document}